\documentclass[11pt,a4paper]{article}
\usepackage{amsmath, amssymb}
\usepackage{fixmath}
\usepackage{hyperref}

\usepackage{geometry}
\usepackage[osf,noBBpl,sc]{mathpazo}

\newtheorem{theorem}{Theorem}

\newcommand\soft[1]{\textsc{#1}}
\DeclareMathOperator\val{val}
\DeclareMathOperator\mult{mult}
\newcommand\QQ{\mathbb Q}
\newcommand\XX{\mathbold{X}}
\renewcommand\aa{\mathbold{\alpha}}
\newcommand\dd{\mathbold{\delta}}

\usepackage{algorithm}
\usepackage[noend]{algpseudocode}
\algrenewcommand\algorithmicrequire{\textbf{Input:}}
\algrenewcommand\algorithmicensure{\textbf{Output:}}
\newcommand\Input\Require
\newcommand\Output\Ensure
\newcommand\false{\textbf{false}}

\begin{document}

\title{\soft{Lacunaryx}:\\Computing bounded-degree factors of lacunary polynomials}

\author{Bruno Grenet\thanks{Partially supported by a LIX--Qualcomm\textsuperscript\textregistered--Carnot postdoctoral fellowship, and by the French Agence Nationale de la Recherche under grant CATREL \#ANR-12-BS02-001.}\\
LIRMM -- UMR 5506 CNRS \\ 
Universit\'e de Montpellier \\
\url{bruno.grenet@lirmm.fr}}

\date{}

\maketitle

\section{Introduction}
In this paper, we report on an implementation in the free software \soft{Mathemagix}~\cite{mmx} of lacunary factorization algorithms, distributed as a library called \soft{Lacunaryx}.\footnote{This library and its documentation can be found at \url{http://mathemagix.org/www/lacunaryx/doc/html/index.en.html}.}
These algorithms take as input a polynomial in sparse representation, that is as a list of nonzero monomials, and an integer $d$, and compute its degree-$d$ factors.\footnote{Here and henceforth, the expression \emph{degree-$d$ factor} means an irreducible factor of degree \emph{at most} $d$.} The complexity of these algorithms is polynomial in the \emph{sparse size} of the input polynomial and $d$: The sparse size of a polynomial $f=\sum_{j=1}^k c_j X_1^{\alpha_{1,j}}\dotsb X_n^{\alpha_{n,j}}\in\QQ[X_1,\dotsc,X_n]$ is $\sum_j(h(c_j)+\sum_i\log(1+\alpha_{i,j}))$ where the $h(\cdot)$ denotes the \emph{height} of a rational number, that is the maximum of the absolute values of its numerator and denominator. 

In our implementation, we focus on polynomials with rational or integer coefficients. Algorithms in the univariate case have been given by Cucker, Koiran and Smale~\cite{CuKoiSma99} and generalized by Lenstra~\cite{Len99}. In the multivariate case, the first algorithms were given by Kaltofen and Koiran~\cite{KaKoi05,KaKoi06}. We described simpler algorithms for multilinear factors, with a different approach, with Chattopadhyay, Koiran, Portier and Strozecki~\cite{ChaGreKoiPoStr13,ChaGreKoiPoStr14} and then generalized them to bounded-degree factors~\cite{Gre14,Gre15}. We implement a variant of Lenstra's algorithm for the univariate case, and our most recent algorithms for the multivariate case. The algorithms of Koiran and Kaltofen are not easily implementable since they rely on the value on some non-explicit number-theoretic constant.

\section{Algorithms}

This section is devoted to the description of the algorithms we implemented. We refer to the original papers for more formal descriptions, proofs of correctness and complexity estimates.

\subsection{Lenstra's algorithm}

Lenstra's algorithm has two disjoint steps for computing \emph{cyclotomic factors} on the one hand, that is factors that divide $X^r-1$ for some $r$, and \emph{non-cyclotomic factors} on the other hand. Let us first describe the step for non-cyclotomic factors.

\begin{theorem}[Lenstra's Gap Theorem]
Let $d$ be a positive integer and $f=f_1+f_2\in\QQ[X]$. There exists a constant $\gamma(f,d)$ such that if $\val(f_2) - \deg(f_1) > \gamma(f,d)$, every non-cyclotomic factor of degree at most $d$ of $f$ divides both $f_1$ and $f_2$.
\end{theorem}

The constant in the theorem is polynomial in the sparse size of $f$ and in $d$.
It yields an algorithm to compute the degree-$d$ non-cyclotomic factors of a given $f\in\QQ[X]$ in time polynomial in the sparse size of $f$ and $d$: 
1. Compute $\gamma(f,d)$ and $f_1$ of minimal degree such that $\val(f-f_1) - \deg(f_1)>\gamma(f,d)$; 2. Recursively compute $f_2$, \ldots, $f_s$ such that $f=f_1+\dotsb+f_s$ with the same property; 3. Compute the factors of $\gcd(f_1,\dotsc,f_s)$. The theorem ensures that every degree-$d$ non-cyclotomic factor of $f$ is a common factor of $f_1$, \ldots, $f_s$, that is a factor of $\gcd(f_1,\dotsc,f_s)$. To compute the multiplicities, the same algorithm is applied to the derivatives of $f$, or more precisely to its sparse derivatives: The sparse derivative $f^{[1]}$ of $f$ is defined as $(f/X^{\val(f)})'$ where $'$ denotes the standard derivative.

The computation of the cyclotomic degree-$d$ factors of $f$ works as follows: 
If $g$ is cyclotomic, it divides by definition $X^r-1$ for some $r$. The cyclotomic factors of $f$ that divide a fixed $X^r-1$ also divide $\gcd(f, X^r-1)$. Let us write $f=\sum_{j=1}^k c_j X^{\alpha_j}$ and define $f^{\bmod r} = \sum_{j=1}^k c_j X^{\alpha_j\bmod r}$. Then $f^{\bmod r}$ satisfies $\gcd(f, X^r-1)=\gcd(f^{\bmod r}, X^r-1)$. This shows that for each $r$, one can compute the common factors of $f$ and $X^r-1$ in polynomial time in $r$ and the sparse size of $f$. 
Number-theoretic considerations show that all degree-$d$ cyclotomic polynomials divide some $X^r-1$ with $r\le 2d^2$.
Altogether, the degree-$d$ cyclotomic factors of $f$ can be computed in polynomial time in $d$ and the sparse size of $f$. Again, the multiplicities are computed using the same algorithm on the sparse derivatives of $f$.

\subsection{Variant of Lenstra's algorithm}

To describe our variant of Lenstra's algorithm, let us reformulate it slightly. With the notation of the previous section, let $g=\gcd(f_1,\dotsc,f_s)$, and $h=f/g$. 
Lenstra's algorithm can be described as follows: 1. Compute $g$ and $h$; 2. Factor $g$; 3. Compute the cyclotomic factors of $h$; 4. Return the union of these factors (viewed as multisets, to take multiplicities into account); 5. Apply the same algorithm to the sparse derivatives of $f$.

We build on this view in our algorithm. Instead of computing a single $g$ and a single $h$, we actually compute two sets $G$ and $H$ such that for any degree-$d$ polynomial $\ell$, 
\begin{equation}\label{eq:GH}
\mult_\ell(f)=
\begin{cases}
\sum_{g\in G}\mult_\ell(g)+\sum_{h\in H}\mult_\ell(h) & \text{if $\ell$ is cyclotomic, and}\\
\sum_{g\in G}\mult_\ell(g) &\text{otherwise}
\end{cases}
\end{equation}
where $\mult_\ell(f)$ denotes the multiplicity of $\ell$ as a factor of $f$.
Let us remark that the most expensive part of Lenstra's algorithm is the factorization of the polynomial $\gcd(f_1,\dotsc, f_s)$. Our modification allows us to reduce as much as possible the degree of the polynomials that have to be eventually factored. 

To compute the sets $G$ and $H$, we reverse the logic of Lenstra's algorithm: While Lenstra computes an \emph{a priori} bound $\gamma(f,d)$ to split $f$ as a sum, we split $f$ as much as possible (say as a sum of monomials) and then merge together the summands, beginning with the closest ones. At each step, if the summands have a non trivial gcd $g$, we apply recursively our algorithm to $f/g$ to get the two sets $G$ and $H$, and return $(G\cup\{g\}, H)$. Else, we merge the two closest summands (that is we replace both of them by their sum) and compute again the gcd of the new summands, until there remains a unique summand. As described, the algorithm would always return an empty set $H$ since the gcd of the summands is computed at each step. Actually, this is useless in the following case: If two summands $f_i$ and $f_{i+1}$ have been merged but $\val(f_{i+1})-\deg(f_i) > \gamma(f_i+f_{i+1},d)$, Lenstra's Gap Theorem ensures that the non-cyclotomic degree-$d$ factors of $f_i+f_{i+1}$ are common to $f_i$ and $f_{i+1}$. The gcd does not need to be computed, and we can directly merge the next summands. When it remains a unique summand, it goes into $H$ rather than $G$ if the gap that has been merged is large enough. The formal description of this algorithm is given as Algorithm~\ref{algo:separation}. Note that the sets $G$ and $H$ give a \emph{partial factorization} of $f$ in the sense that $f = \prod_{\ell\in G\cup H} \ell$.

\begin{algorithm}
\caption{Bounded-degree partial factorization of $f$} 
\label{algo:separation}
\begin{algorithmic}[1]
\Input $f=\sum_{j=1}^k c_j X^{\alpha_j}$ with $\alpha_1\le\dotsb\le\alpha_k$, and $d>0$;
\Output The sets $G$ and $H$ as defined in Eq.~\eqref{eq:GH}.
\Statex
\Procedure{Partial\_factorization}{$f, d$}
\State $S\gets [c_1X^{\alpha_1}, \dotsc, c_kX^{\alpha_k}]$
\State $b\gets\false$ \Comment{Only monomials}
\While{$|S|>1$}
    \If{$b$}
        \State $g\gets\gcd(S)$
        \If{$\deg (g) > 0$} 
            \State $(G,H)\gets\textsc{Partial\_factorization}(f/g, d)$
            \State \textbf{return} $(G\cup\{g\}, H)$
        \EndIf
    \EndIf
    \State $t\gets$ index that minimizes $\delta=\val(S[t+1]) - \deg(S[t])$
    \State $b\gets (\delta>\gamma(S[t]+S[t+1],d))$ \Comment{\textbf{true} or \textbf{false}}
    \State $S[t]\gets S[t]+S[t+1]$
    \State $S\gets S\setminus\{S[t+1]\}$ 
\EndWhile
\If{$b$} \textbf{return} $(\{S[0]\}, \emptyset)$
\Else\ \textbf{return} $(\emptyset, \{S[0]\})$
\EndIf
\EndProcedure
\end{algorithmic}
\end{algorithm}

After $G$ and $H$ have been computed, it remains to compute all the factors of the polynomials in $G$ using any classical algorithm, and the cyclotomic ones for polynomials in $H$ using a variant of Lenstra's algorithm for cyclotomic factors. Finally, the union of these factors is returned. We remark that our method directly provides the multiplicities of the non-cyclotomic factors.

Let us describe our variant of Lenstra's algorithm for cyclotomic factors.
Let $h\in H$ be a polynomial, and suppose we aim to compute its degree-$d$ cyclotomic factors.
As mentioned in the previous section, a cyclotomic factor divides $X^r-1$ for some $r$. Lenstra proved that $r\le2d^2$ for a degree-$d$ cyclotomic factor. 
Actually, one can compute a much smaller set $R\subsetneq\{1,\dotsc, 2d^2\}$ such that each degree-$d$ cyclotomic polynomial divides $X^r-1$ for some $r\in R$. 
By definition, the $r$-th cyclotomic polynomial, denoted by $\phi_r$, is the only irreducible polynomial that divides $X^r-1$ and does not divide any $X^s-1$ for $s<r$. Furthermore, it is known that $\deg(\phi_r)=\varphi(r)$ where $\varphi$ is Euler's totient function. That is, the cyclotomic degree-$d$ factors of $h$ divide some $X^r-1$ with $\varphi(r)\le d$. In our implementation, we compute the set $R=\{r\le 2d^2:\varphi(r)\le d\}$ in a naive way, 
and for each $r\in R$, we exactly compute the $r$-th cyclotomic polynomial as $\phi_r = (X^r-1) / \gcd(X^r-1, \prod_{s\in R, s<r} \phi_s)$. Thus, instead of computing the factors of $\gcd(h,X^r-1)$ as in Lenstra's original algorithm, we compute $\phi_r$ and simply test whether it divides $h$, using again the fact that $\phi_r$ divides $h$ if and only if it divides $h^{\bmod r}$.

\subsection{Multivariate case}

The multivariate case has the same high-level structure as the univariate case. Namely, the computation of the \emph{unidimensional factors}, that is multivariate factors that can be written as $f(\XX)=\XX^\aa f_\pi(\XX^\dd)$ where $\XX^\aa$ and $\XX^\dd$ are multivariate monomials and $f_\pi$ is a univariate polynomial of valuation $0$, reduces to the factorization of some univariate polynomials, while the computation of the \emph{multidimensional factors} is based on a \emph{Gap Theorem}.  

The algorithms we implement for these two steps are very close to the one described in our previous works~\cite{Gre14,Gre15}. For the multidimensional factors, we apply the same modification as for Lenstra's algorithm in the univariate case, that is we first write $f$ as a sum of monomials and then iteratively merge the summands. This way, we actually build a so-called \emph{single-linkage clustering} of the set of exponents.

\section{Examples}

We include an example of timings that is pretty representative of the computation times we obtain. It is not comparable to other softwares since in the range of degrees that we consider, other software are unable to give any answer. (Often, they even do not accept large enough exponents for polynomials.) Also, the nature of the algorithms we use make the computation times very versatile: A very small change in the input polynomial may force the algorithm to factor a much larger polynomial, or to compute a gcd with much larger inputs. This raises two important questions: Can we (automatically) provide an estimation of the size of the polynomials that will appear in the computations? Can we modify the algorithms to reduce their versatility, or implement strategies to compute factors of degree as large as possible while guaranteeing \emph{reasonable} computation times? 

We consider a random polynomial of degree $>1\,000\,000$ and $>10\,000$ nonzero terms constructed as a product of three parts: a product of $5$ degree-$10$ polynomials, a product of $3$ polynomials of the form $X^r-1$ with $r$ of order $100\,000$ and a sparse polynomial of degree $1\,000\,000$ with $40$ monomials each multiplied by polynomials of degree $20$. Next table gives the timings to compute degree-$5$, $10$, $15$, and $20$ factors. We observe that the main part of the computation rapidly becomes the gcd computations.

The code for this example is distributed as part of the \soft{Lacunaryx} package, in the file \texttt{lacunaryx/bench/bounded\_degree\_univariate\_bench.cpp}. The timings were obtained on an \emph{Intel}\textcopyright{} \emph{Core}\texttrademark{} \emph{CPU @ 2.60GHz} platform with \emph{7.7GB RAM}.

\begin{center}
\begin{tabular}{|l|c|c|c|c|}
\hline
Degree of the factors & $5$  & $10$ & $15$  & $20$ \\\hline
Total time (ms)       & 1994 & 2924 & 12190 & 26165 \\
Non-cyclotomic (ms)   & 1948 & 2856 & 12140 & 26061 \\
Cyclotomic (ms)       & 46   & 68   & 50    & 104 \\
Gcd computations (ms) & 91   & 749  & 9223  & 23151 \\
\hline
\end{tabular}
\end{center}

\end{document}